\documentclass[aip,amsmath,amssymb,reprint,groupedaddress,floatfix]{revtex4-1}

\usepackage{graphicx}
\usepackage{bm}

\usepackage{amsmath}
\usepackage{amssymb}
\usepackage{color}
\usepackage{subfigure}
\usepackage{times}
\usepackage{float}
\usepackage[sort&compress]{natbib}

\usepackage[colorinlistoftodos,prependcaption,textsize=small]{todonotes}

\usepackage{hyperref}

\hypersetup{
  colorlinks=true,
citecolor=black,
linkcolor=black,
urlcolor=black
  }

\usepackage[T1]{fontenc}
\usepackage[utf8]{inputenc}

\newcommand{\egn}[1]{{\color{black} #1}}
\newcommand{\bdt}[1]{{\color{blue} #1}}

\renewcommand{\bdt}[1]{{\color{black} #1}}

\newcommand{\xr}{x_{\mathrm{res}}}

\newcommand{\kct}[1]{\textcolor{blue}{#1}}
\renewcommand{\kct}[1]{\textcolor{black}{#1}}

\begin{document}

\title{Dichotomous flow with thermal diffusion and stochastic resetting}
\author{Karol Capa{\l}a}
\email{karol@th.if.uj.edu.pl}

\author{Bart{\l}omiej Dybiec}
\email{bartek@th.if.uj.edu.pl}

\author{Ewa Gudowska-Nowak}
\email{ewa.gudowska-nowak@uj.edu.pl}

\affiliation{Institute of Theoretical Physics, Department of Statistical Physics, Jagiellonian University, \L{}ojasiewicza 11, 30-348 Krak\'ow, Poland}


\date{\today}

\begin{abstract}
We consider properties of one-dimensional diffusive dichotomous flow and discuss effects of \bdt{stochastic} resonant activation (SRA) in the presence of statistically independent random resetting mechanism.
Resonant activation and stochastic resetting are two similar effects, as both of them can optimize the noise induced escape.
Our studies show completely different origins of optimization in adapted setups.
Efficiency of stochastic resetting relies on elimination of suboptimal trajectories, while SRA is associated with matching of time scales in the dynamic environment.
Consequently, both effects can be easily tracked by studying their asymptotic properties.
Finally, \egn{we show that} stochastic resetting cannot be easily used to further optimization of the  SRA  in symmetric setups.
\end{abstract}


\maketitle

\setlength{\tabcolsep}{0pt}

\textbf{
 \bdt{Stochastic} resonant activation (SRA) is a celebrated, well known and extensively studied effect demonstrating that combined action of the barrier modulation process and noise can result in the optimal, as measured by the mean first passage time, escape kinetics.
The efficiency of the SRA relies on matching of time scales characterizing noise driven escape and potential modulation.
Another renowned effect that can be used to facilitate noise induced escape is stochastic resetting.
Within stochastic resetting the increase in the escape rate is achieved by elimination of suboptimal meandering trajectories.
Here, we combine and compare both methods of facilitating escape rates showing their similarities and inherent differences.
Using numerical and analytical methods, we study properties of both effects \egn{and demonstrate} that typically applied criteria for determining the area in which resetting enhances escape kinetics should be used with special care in the regime \egn{of dynamic modulation of the potential}.
}

%
%
\section{Introduction and model}

Typically, noise observed in electronic devices or in communication systems is assumed detrimental. Astounding, although it is an inevitable component of their structure and dynamics, under certain circumstances, it can be beneficial in detection of weak signals and system synchronization.
The action of noise can result in occurrence of various constructive effects like \bdt{dynamical resonant activation \cite{devoret1984}}, stochastic resonance \cite{mcnamara1989,gammaitoni1998,spagnolo2002role,mantegna2000linear}, \egn{SRA} \cite{doering1992} or noise-induced stabilization\cite{dubkov2003nonequilibrium,spagnolo2004,dubkov2007,fiasconaro2006co,fiasconaro2008monitoring}, to name just a few.
\egn{These beneficial noise induced effect can be applied, among others, in technical \cite{dubkov2005,falci2013design}, economic \cite{spagnolo2008volatility}, biological \cite{russell1999use,ward2002stochastic}, geological \cite{wiesenfeld1995} and medical realms \cite{priplata2003vibrating,kaut2011stochastic}.}
One of the paradigmatic examples is action of noise in transport phenomena and \egn{fluctuation-driven} crossing over the potential barrier \cite{redner2001}. The latter can be further optimized by the additional time-dependent modulation \egn{(either deterministic, periodic or stochastic)} of the potential barrier \cite{doering1992,pechukas1994rates,pankratov2000}, leading to the emergence of resonant activation. 
Another mechanism which can optimize the noise induced escape is stochastic resetting \cite{evans2011diffusion,evans2011diffusion-jpa,evans2020stochastic}.
The fact that these two very different in nature processes can optimize the escape rate calls for investigation of combined action of the potential modulation and stochastic resetting.
This in turn \egn{should} shed light on similarities and differences between them.

%
%
\begin{figure}[!h]%
\begin{center}
\includegraphics[angle=0,width=0.95\columnwidth]{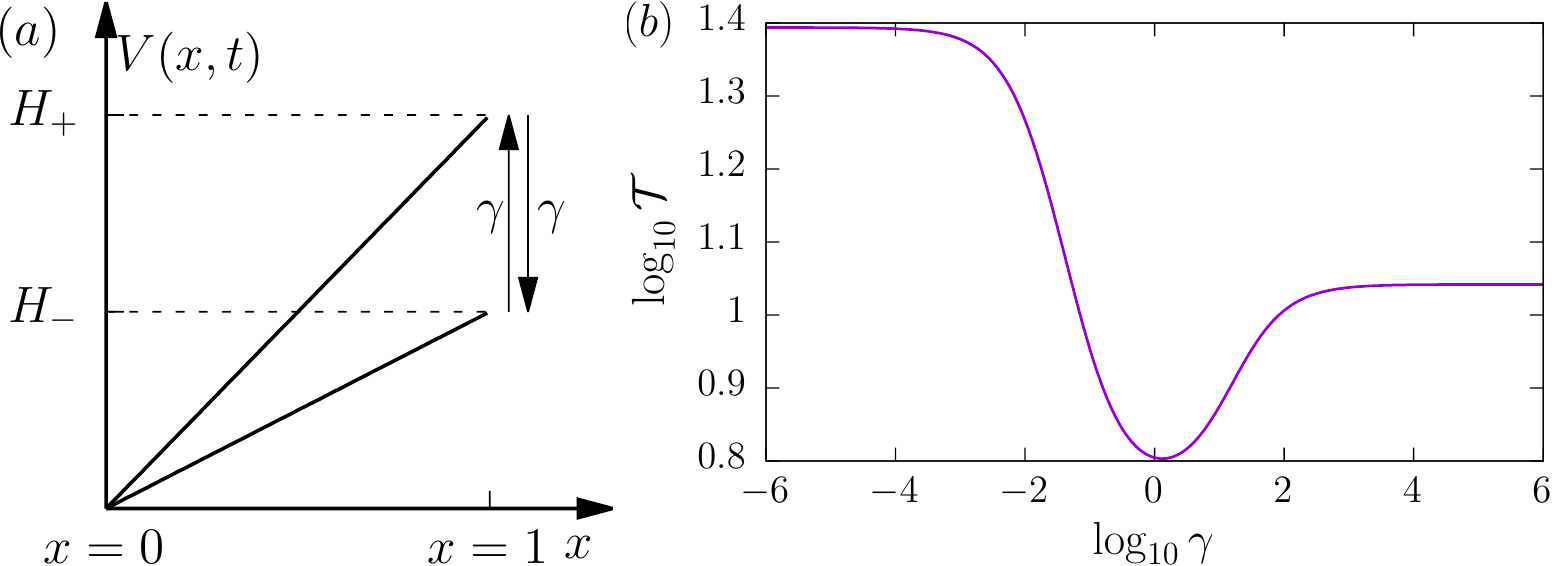}
\caption{The classical SRA setup (left panel --- ($a$)) the dichotomously switching linear potential between two configurations characterized by distinct heights $H_\pm$ and the typical SRA curve (right panel --- ($b$)) for $H_+=8$ and $H_-=4$ with $x(0)=0$ showing dependence of the MFPT on the switching rate $\gamma$.}
\label{fig:setup}
\end{center}
\end{figure}

In the resonant activation phenomenon, analogously to the stochastic resetting \cite{pal2019first} there is the optimal barrier modulation protocol resulting in the shortest mean escape time.
The phenomenon can be \egn{analyzed} in models described by the following overdamped Langevin equation
\begin{equation}
  \frac{dX}{dt}=-V'(X,t)+\sqrt{2T}\xi(t),
  \label{eq:langevin}
\end{equation}
where $-V'(x,t)= -dV(x,t)/dx$ \bdt{is the time dependent force and $\xi(t)$ is the Gaussian white noise satisfying $\langle \xi(t) \rangle=0$ with $\langle \xi(t)\xi(s) \rangle = \delta(t-s)$.} 
\egn{Here $\sqrt{2T}$, where $T$ stands for a fixed system temperature, gives the noise intensity which controls the strength of fluctuations. Within simulations we further assume $T=1$ with the friction constant incorporated in the time scale.}
\bdt{The SRA phenomenon \cite{doering1992}, which is studied here,  should be distinguished from the dynamical resonant activation \cite{devoret1984} (DRA)
 observed e.g. in Josephson junction, when the external driving frequency is close to the natural characteristic frequency of the system \cite{devoret1985measurements,martinis1987experimental}.
In contrast, the SRA relies on matching of time scales characterizing potential modulation and the noise driven escape \cite{pan2009resonant,valenti2014,guarcello2015phase}, which is controlled by the system temperature.
}

The seminal \egn{SRA} setup was suggested in Ref.~\onlinecite{doering1992}, where the symmetric, Markovian, dichotomous modulation \cite{horsthemke1984} of the linear potential was considered.
As a result, the potential was switching \bdt{(with the flipping rate $\gamma$)} between two slopes of various height $H_\pm$:
$
 V(x,t)=V_\pm(x)=H_\pm x.
$
The deterministic force determined by the linear potential introduces dichotomous flow which is affected by presence of boundaries and subjected to additional thermal diffusion.
Therefore, the studied model goes beyond the telegraphic processes with stochastic resetting studied in Ref.~\onlinecite{masoliver2019telegraphic} and models of run-and-tumble particles with resetting, examined in Refs.~\onlinecite{bressloff2020,evans2018}.
The overall process with the potential switching between two configurations is then described by the Chapman-Kolmogorov equation
\begin{eqnarray}
\frac{\partial}{\partial t}
\left[
    \begin{array}{c}
         P_+  \\
         P_-  
    \end{array}
    \right]
     & = &
    \left[
    \begin{array}{cc}
         -H_++T \frac{\partial^2}{\partial x^2}  & 0\\
         0 &  -H_- +T\frac{\partial^2}{\partial x^2}
    \end{array}
    \right]
    \left[
        \begin{array}{c}
         P_+  \\
         P_-
    \end{array}
    \right] \nonumber \\
    & + &
    \left[
        \begin{array}{cc}
         -\gamma & \gamma  \\
         \gamma & -\gamma
    \end{array}
    \right]
    \left[
        \begin{array}{c}
         P_+  \\
         P_-
    \end{array}
    \right],
    \label{eq:kolmogorov}
\end{eqnarray}
where $P_\pm$ is the probability density of the particle at the position $x$ at time $t$ moving along the potential slope in one of possible ($\pm$) configurations.
In Eq.~(\ref{eq:kolmogorov}), the first term on the right hand side describes the diffusion in the external potential, while the second one switching of the potential.
\bdt{On average, the barrier stays in one of configurations before switching to another for time equal to $1/\gamma$.}
Initially any configuration of the potential is equally probable.
Moreover, the domain of motion is restricted by two boundaries: reflecting boundary at $x=0$ and absorbing one at $x=1$, see Fig.~\ref{fig:setup}($a$).
In the SRA phenomenon the quantity of interest is the mean first passage time (MFPT) $\mathcal{T}$ which is the average of first passage times $t_{\mathrm{fp}}$
\begin{equation}
    \mathcal{T} =   \langle t_{\mathrm{fp}} \rangle =
     \langle \min\{t : x(0)=0 \;\land\; x(t) \geqslant 1  \} \rangle.
\end{equation}
In the SRA, the MFPT is a non-monotonous function \cite{doering1992} of the barrier modulation parameter, e.g., $\gamma$ --- the transition rate of the dichotomous noise. There exists such a value of $\gamma$ for which the MFPT is minimal, see Fig.~\ref{fig:setup}($b$).
The minimal MFPT is recorded when the majority of escapes is performed over the lowest barrier configuration \cite{doering1992,pechukas1994rates,iwaniszewski2003,fiasconaro2011resonant}.
It demonstrates that efficiency of SRA relies on matching of time scales associated with the modulation of the potential barrier and escape time over a fixed potential slope.
Moreover, two special limiting cases \cite{doering1992,iwaniszewski1996} are recorded:   the MFPT is equal to the average of MFPT over possible barrier configurations ($\gamma\to 0$) or it is equal to the MFPT over average barrier ($\gamma\to \infty$).
Such asymptotics are recorded because for large $\gamma$ a particle feels the average potential while in the $\gamma \to 0$ limit the escape takes place over a fixed, randomly selected, potential barrier.

Here, we extend the classical SRA model by incorporation of the stochastic resetting \cite{evans2011diffusion,evans2011diffusion-jpa,evans2020stochastic}.
We assume that at random time instants $t_i$ the particle position is reset by putting it back to the origin.
The time intervals between two consecutive resets (inter-resetting times) follow the exponential distribution.
The method of \bdt{Poissonian} resetting is described and explored in the next section (Sec.~\ref{sec:results}).
The manuscript is closed with Summary and Conclusions (Sec.~\ref{sec:summary}).
More detailed calculations are moved to Appendix.

\section{Stochastic resonant activation with Poissonian resettings\label{sec:results}}
To this end we define the stochastic process of diffusion with Poissonian resettings. The underlying diffusion process $\{X(t)\}$ starts at $x=0$ and evolves over a certain (random) time on the interval $[0, 1]$ with a linear drift $H_{\pm}$. At the end of this time, the process is reset to some position $\xr$ which can be chosen arbitrarily, e.g., $\xr=0$, and restarts its motion.
We assume that the resets are Poisson-like events occurring in time with exponential distribution of time duration $\tau$ between two subsequent resets, $\phi(\tau)=r \exp{(-r \tau)}$, where $r$ is the resetting rate. By assuming that the underlying stochastic dynamics is statistically independent from the resetting, the simple renewal equations for the overall process can be derived \cite{Sokolov2019,evans2020stochastic} and some observables (like the survival probability of not crossing the barrier up to given time) in the presence of resetting may be expressed in terms of the same observables for a process free from resets, see Ref.~\onlinecite{reuveni2016optimal} and~Appendix.

\begin{figure}[h]%
\centering
\includegraphics[angle=0,width=0.95\columnwidth]{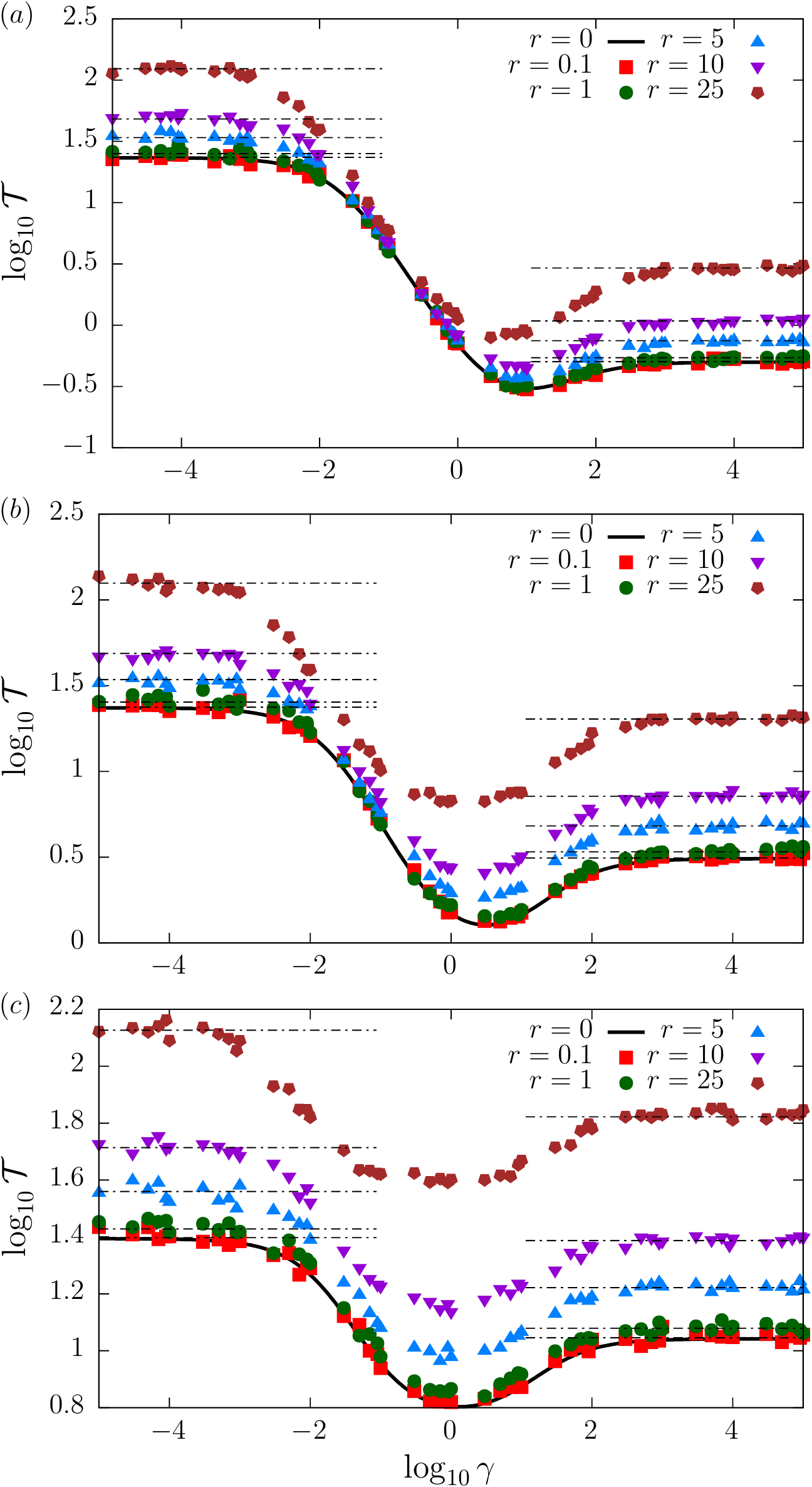} 
\caption{MFPT as a function of the barrier switching rate $\gamma$ for $H_{\pm}=\pm 8$ (top panel --- ($a$)), $H_{\pm}=8/0$ (middle panel --- ($b$)) and $H_{\pm}=8/4$ (bottom panel --- ($c$)) \bdt{with $T=1$}.
Different curves correspond to various resetting rates $r$.
Black thin dot-dashed lines present average values of MFPTs over both barrier configurations ($\gamma\to 0$) and MFPTs over the average barrier ($\gamma \to \infty$) calculated by use of Eqs.~(\ref{eq:ra-free-large-gamma}) and~(\ref{eq:mfpt-well-reset}).
}
\label{fig:ra-reset}
\end{figure}

Sample MFPT curves corresponding to various configurations of the potential barrier $H_\pm x$ with \bdt{$T=1$ and} different reset rates $r$ are presented in Fig.~\ref{fig:ra-reset}.
As exemplary barrier heights we have selected the traditionally considered, see Ref.~\onlinecite{doering1992},  $H_\pm=\pm 8$ (Fig.~\ref{fig:ra-reset}($a$)),
$H_\pm = 8/0$ (Fig.~\ref{fig:ra-reset}($b$)) and $H_\pm=8/4$ (Fig.~\ref{fig:ra-reset}($c$)).
MFPT curves have been estimated from an ensemble of first passage times obtained by stochastic simulation the Langevin equation (\ref{eq:langevin}) with the Euler-Maruyama method \cite{higham2001algorithmic,mannella2002} with $\Delta t=10^{-5}$ and averaged over $N=10^3$ realizations.
Various curves correspond to various values of reset rate $r$ and consequently various average interresetting times $\langle \tau \rangle=1/r$.
In all cases under study obtained curves follow the typical, non-monotonous, SRA curve shape, see Ref.~\onlinecite{doering1992} and Fig.~\ref{fig:setup}($b$).
Solid lines present exact values of the MFPT \cite{mus} for the SRA without resetting, i.e., for $r=0$ ($\langle \tau \rangle = \infty$).
For $r=0.01$ ($\langle \tau \rangle = 100$) resets are performed so rarely that obtained results are indistinguishable from the without resetting case (results not shown), while for $r=0.1$ they are very close to the without resetting case, see Fig.~\ref{fig:ra-reset}.
With the increasing reset rate $r$, i.e., decreasing average interresseting time $\langle \tau \rangle$, MFPTs increase, minima become deeper and shifted towards smaller frequencies $\gamma$.
For all SRA curves low frequency $\lim_{\gamma\to0} \mathcal{T} = [\mathcal{T}(H_-)+\mathcal{T}(H_+)]/2$
and high frequency $\lim_{\gamma\to\infty} \mathcal{T} = \mathcal{T}( (H_-+H_+)/2)$ limits are perfectly recovered, see thin dot-dashed lines in Fig.~\ref{fig:ra-reset} and Appendix.
In other words, in the limit of $\gamma\to 0$ the MFPT is equal to the average of MFPTs over both barrier configurations, while in the limit of $\gamma\to\infty$ the MFPT is equal to the MFPT over the average potential barrier.

%
%

The increase in the resetting rate $r$ results in the increased MFPTs, see Fig.~\ref{fig:ra-reset}.
The resetting is responsible for multiple returns to the origin, i.e., $\xr=0$, and consequently for the increase in the MFPT as the $x=0$ is the point corresponding to the statistically slowest escape.
Nevertheless, one can naively think that the increasing MFPT in Fig.~\ref{fig:ra-reset} is related to the decrease in the system temperature as typically for the decreasing \bdt{temperature} $T$ the MFPT increases \cite{gardiner2009,doering1992}.
If it were the case, one could fit the results for given $r$ by decreasing the system temperature, which definitely is not the case.
This can be well demonstrated by studying the escape from finite intervals.
If the changes in the system temperature could compensate resetting the similar behavior would be visible for the escape from the finite interval $[-L,L]$ restricted by two absorbing boundaries under stochastic resetting when the MFPT \cite{pal2019first} is given by
\begin{equation}
\mathcal{T}(x_0) = \frac{1}{r} \left[ \frac{  \sinh  \frac{2L}{\sqrt{  T/r}}   }{    \sinh \frac{L-x_0}{\sqrt{ T/r}}   +\sinh   \frac{x_0+L}{\sqrt{  T/r}}  } -1 \right].
    \label{eq:mfptreset}
\end{equation}
As it is clearly visible from Eq.~(\ref{eq:mfptreset}), due to the $1/r$ in front of the square bracket, the average interreset time, $\langle \tau \rangle = 1/r$, cannot be incorporated into the system temperature $T$.
Consequently, also in the case of SRA with restarting, one cannot combine resetting and temperature.
Interestingly, both the increase of the resetting rate $r$ or the decrease in the system temperature $T$, increases the mean first passage time.
However, the mechanisms of increase in the MFPT are fundamentally different.
\bdt{First of all, it is visible on the level of Eq.~(\ref{eq:mfptreset}) which shows that the MFPT under resetting is not a function of $r T^\beta$. Therefore, the action of resetting cannot be compensated by adjusting the system temperature.
Secondly, it can be intuitively demonstrated:
}
Decrease in system temperature weakens fluctuations and hardens surmounting of the potential barrier, while the increase in the reset rate increases the number of revisits to $\xr=0$ and increases the amount of unsuccessful attempts to overcome the potential barrier.
Moreover, forced returns have to increase the MFPT as $x=0$ is \bdt{is the most distant point from the absorbing boundary.
Therefore, it is the point with the largest MFPT to the target (positioned at $x=L$).}

\bdt{Difference between resetting and decreasing temperature $T$ is most striking for a free particle.
Without resetting position $x(t)$ follows the normal distribution with the zero mean and the linearly growing variance, i.e., $\sigma^2(t)=2T t$.
In the presence of resetting, it attains nonequilibrium stationary density given by the Laplace distribution \cite{evans2020stochastic}.}
\egn{}

Equation~(\ref{eq:mfptreset}) can be not only used to show different role of the temperature and resetting, but also to estimate large $\gamma$ asymptotics of MFPTs in Fig.~\ref{fig:ra-reset}($a$).
In Fig.~\ref{fig:ra-reset}($a$) the potential switches between $\pm H$ configurations, consequently for large $\gamma$ a particle is practically free, i.e., $V(x) \approx 0$.
In such a case the SRA setup (reflecting boundary at $x=0$ and absorbing boundary at $x=1$ ) with $x(0)=0$ is equivalent to the escape from the finite interval restricted by two absorbing boundaries placed at $\pm 1$, see Refs.~\onlinecite{zoia2007,dybiec2017levy,gardiner2009}.
Therefore, the MFPT can be calculated from Eq.~(\ref{eq:mfptreset}) with $L=1$ and $x_0=0$ resulting in
\begin{equation}
\mathcal{T}(0)
=
\frac{1}{r}  \left[ \frac{  \sinh  \sqrt{\frac{4r}{T} }   }{    2 \sinh  \sqrt{\frac{r}{T}} }   -1 \right].
\label{eq:ra-free-large-gamma}
\end{equation}
For $T=1$, Eq.~(\ref{eq:ra-free-large-gamma}) gives exactly the large $\gamma$ asymptotic recorded in Fig.~\ref{fig:ra-reset}($a$).
In Fig.~\ref{fig:ra-reset} low and large $\gamma$ asymptotic (except $H=0$ case) can be calculated from the analog of Eq.~(\ref{eq:mfptreset}) for a particle moving in a potential $V(x)=H|x|$ restricted by two absorbing boundaries.
For $T=1$, $L=1$ the formula for MFPT, see Eq.~(\ref{eq:mfpt-pot-reset}), reads
\begin{equation}
\mathcal{T}(0) = \frac{ e^{H   A} \left(A-1\right)-2A   e^{\frac{1}{2} H \left(A-1\right)} +A+1 }{2 r    A e^{\frac{1}{2} H \left(A-1\right)}}
\label{eq:mfpt-well-reset}
\end{equation}
   where
   \begin{equation}
   A=\sqrt{\frac{4 r }{H^2}+1}.
   \end{equation}
Eq.~(\ref{eq:mfpt-well-reset}) is the special case of the formula for the MFPT, see Eq.~(\ref{eq:mfpt-pot-reset}), in the $V(x)=H |x|$ with  two absorbing boundaries located at $\pm L$ which is derived in Appendix following methods applied in Ref.~\onlinecite{singh2020resetting} and Refs.~\onlinecite{saed2019first,ray2020space,ray2020diffusion}.
Therefore, in Fig.~\ref{fig:ra-reset} black thin dot-dashed lines show theoretical values given by Eqs.~(\ref{eq:ra-free-large-gamma}) and~(\ref{eq:mfpt-well-reset}).
Very small discrepancies (typically less than 2\%) between results of simulation and theoretical values, especially for large $r$, can be further reduced by the decreasing the integration time step and increasing the number of repetitions.
Definitely, in the situation when $\xr=0$ resetting can only increase the time needed to pass over the potential barrier.
In order to observe benefits of the stochastic resetting $\xr$ has to be large enough.
On the other hand, it is reasonable to assume that $\xr$ is not too close to the absorbing boundary, because resetting (instead of surmounting the potential barrier) can become the dominating method of transition over the potential barrier.

%
%
\begin{figure}[h]%
\centering
\includegraphics[angle=0,width=0.95\columnwidth]{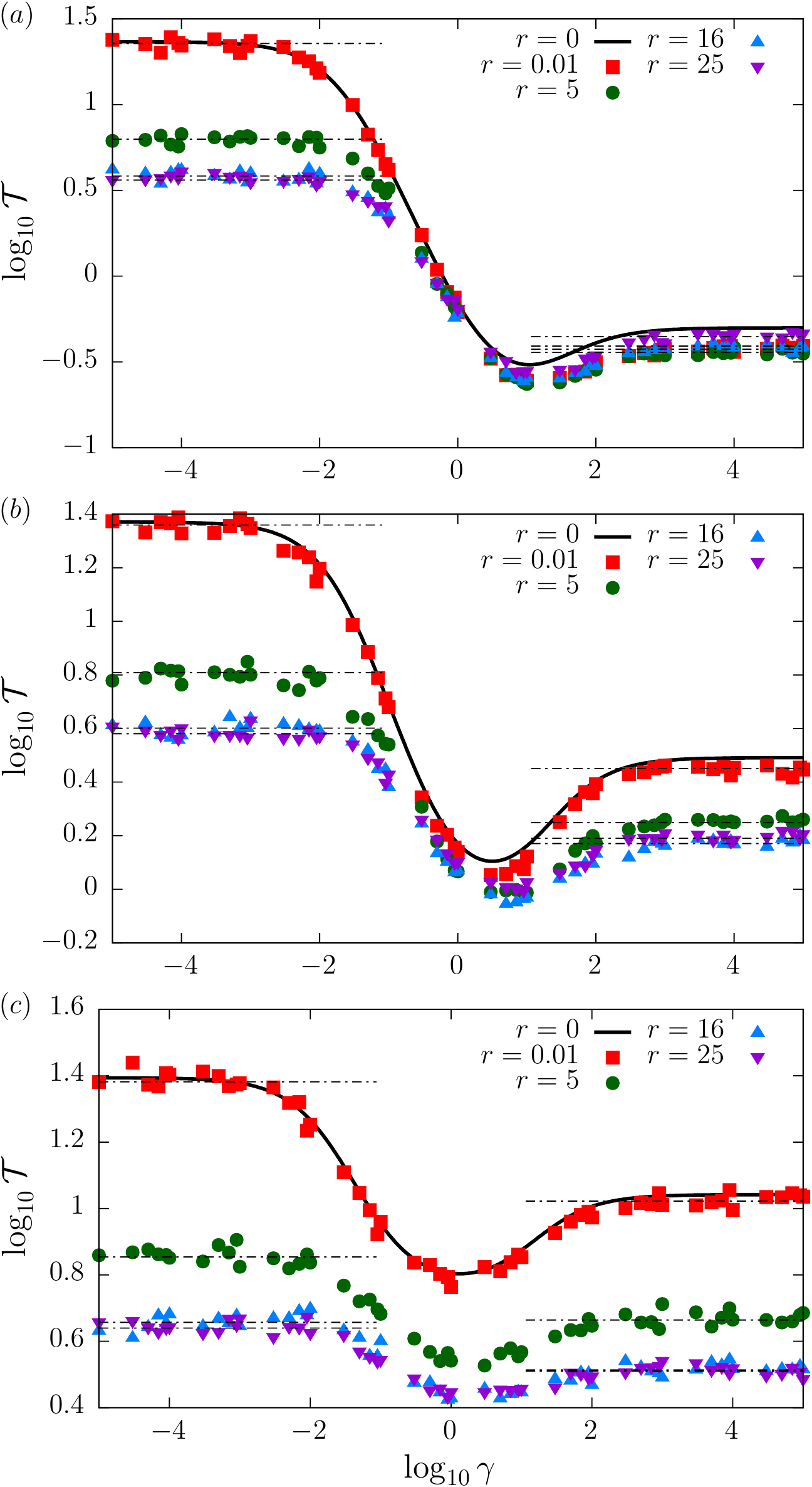} 
\caption{
\bdt{
MFPT as a function of the barrier switching rate $\gamma$ for $H_{\pm}=\pm 8$ (top panel --- ($a$)), $H_{\pm}=8/0$ (middle panel --- ($b$)) and $H_{\pm}=8/4$ (bottom panel --- ($c$)) for $\xr=0.5$ with $T=1$. Different curves correspond to various values of resetting rates $r$.}
Black thin dot-dashed lines show theoretical values calculated with the help of Eqs.~(\ref{eq:mfptreset}) and~(\ref{eq:mfpt-pot-reset}) with appropriately adjusted parameters, while solid lines are the SRA curves for the basic model, see the main text and Fig.~\ref{fig:ra-reset}.
}
\label{fig:ra-reset-x05}
\end{figure}

Figure~\ref{fig:ra-reset-x05} studies the classical \bdt{(stochastic)} resonant activation setup for $\xr=0.5$ with $x(0)=\xr$.
Now, in contrast to Fig.~\ref{fig:ra-reset}, resetting brings a particle to $\xr=0.5$.
Therefore, resetting can be used to facilitate the escape kinetics.
In Fig.~\ref{fig:ra-reset-x05} the black solid lines correspond to the classical without resetting model of SRA with $x(0)=0$, see Ref.~\onlinecite{doering1992} and Fig.~\ref{fig:ra-reset}.
Therefore, results of simulations with $r=0.01$, especially for large $\gamma$ and $H_-=-8$ do not follow solid lines.
The difference \kct{between points and solid line} is caused by the initial condition (solid line $x(0)=0$ versus symbols $x(0)=\xr=0.5$).
For $H_-=0$ or $H_-=4$ the difference is very small because the average deterministic force efficiently moves particles back to the origin making the model with $r=0.01$ very close to the classical SRA setup.
For $r\in\{5,16,25\}$ one can clearly see that resetting further optimizes the SRA, i.e., the resetting can significantly decrease MFPTs.
In the limit of $r\to\infty$ the MFPT diverges, like in all setups with resetting, because the particle is immediately returned to $\xr=0.5$ and it has no chance to reach the absorbing boundary at $x=1$.
For the case with resetting for the low barrier switching rate   $\lim_{\gamma\to0} \mathcal{T} = [\mathcal{T}(H_-)+\mathcal{T}(H_+)]/2$
and high frequency $\lim_{\gamma\to\infty} \mathcal{T} = \mathcal{T}( (H_-+H_+)/2)$ limits are perfectly recovered, see thin dot-dashed lines in Fig.~\ref{fig:ra-reset-x05}.
Low and large $\gamma$ asymptotics have been calculated with the use of MFPTs given by Eqs.~(\ref{eq:mfptreset}) and~(\ref{eq:mfpt-pot-reset}) with $x=\xr=0.5$, $L=1$, $T=1$ and appropriate values of $H$.
In other words, under stochastic resetting with $\xr=0.5$, in the limit of $\gamma\to 0$ the MFPT is equal to the average of MFPTs over both barrier configurations, while in the limit of $\gamma\to\infty$ the MFPT is equal to the MFPT over the average potential barrier.

%
%
\begin{figure}[h]%
\centering
\includegraphics[angle=0,width=0.95\columnwidth]{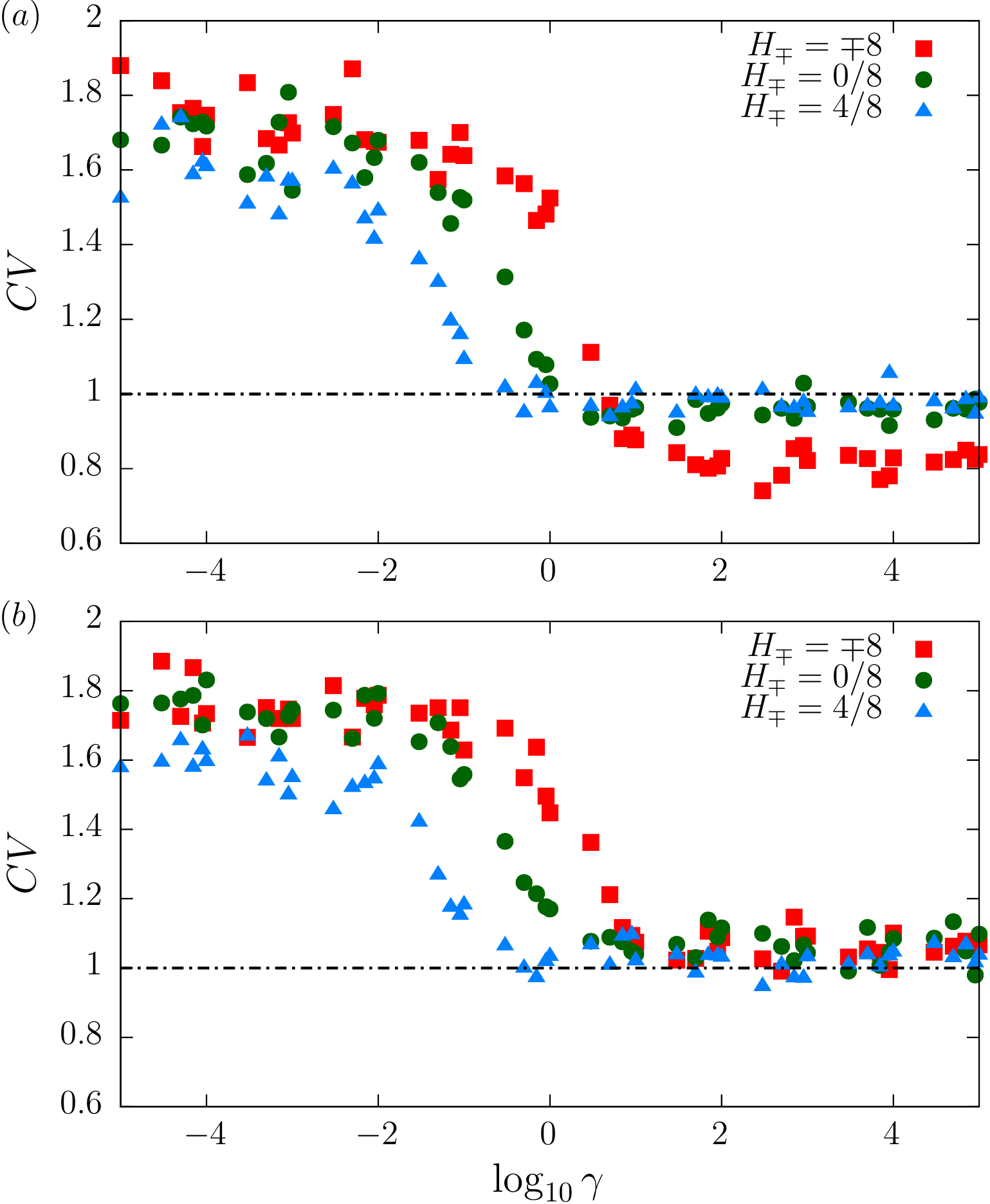} 
\caption{
Coefficient of variation ($CV$) for $\xr=0$ (top panel -- ($a$)) and for  $\xr=0.5$ (bottom panel -- ($b$)).
}
\label{fig:cv}
\end{figure}

Typically, the impact of stochastic resetting is determined by the coefficient of variation (CV)
\begin{equation}
    CV= \frac{\sigma(  t_{\mathrm{fp}})  }{ \langle t_{\mathrm{fp}} \rangle} = \frac{\sigma(  t_{\mathrm{fp}})}{\mathcal{T}},
\end{equation}
which is the ratio between the standard deviation \bdt{$\sigma(t_{\mathrm{fp}})$} of the first passage times and the mean first passage time \bdt{$\mathcal{T}$} in the absence of stochastic resetting \cite{pal2019first}.
\bdt{For the underdamped free motion\cite{pal2019first} or motion in fixed potentials\cite{singh2020resetting}, $CV$ is a convenient tool for evaluating the beneficiality of stochastic resetting.
It is calculated for the system without reset, but $CV$ provides information on what expect in the system with resets.
Using general considerations \cite{pal2017first},
it was shown that for $CV>1$ stochastic resetting can lead to the reduction of the average escape time while for $CV<1$ it slows down the overall escape process.
Here, we explore the usefulness of this measure for motion in fluctuating potentials.
}

Figure~\ref{fig:cv} shows the dependence of the numerically estimated coefficient of variation on the switching rate $\gamma$ corresponding to both setups studied in Fig.~\ref{fig:ra-reset} (top panel -- ($a$)) and Fig.~\ref{fig:ra-reset-x05} (bottom panel -- ($b$)).
When a particle is reset to $\xr=0.5$ the resetting decreases the mean first passage time, see Fig.~\ref{fig:ra-reset-x05}, and numerically estimated coefficient of variation is larger than 1.
The more intriguing situation is recorded for $\xr=0$.
As it is clearly visible from Fig.~\ref{fig:ra-reset} resetting hinders the escape process as it brings a particle back to the point from which the mean first passage time is maximal.
Nevertheless, in this case the coefficient of variation is not always smaller than 1.
The fact that the escape process cannot be optimized by the stochastic resetting in the situation when $CV>1$ is due to additional external modulation of the potential barrier.
This means that in comparison to typical setups, e.g., escape from the finite interval or fixed potential, \bdt{$CV$ can be only used as a proxy for assessing benefit of resetting.
In the studied SRA setup, the resetting is performed in the underdamped system with non-fixed potential.
}
Consequently, the fact that $CV>1$ does not necessarily mean that resetting can accelerate the absorption which is especially well visible for small $\gamma$, see Fig.~\ref{fig:ra-reset}.

%
%
\begin{figure}[h]%
\centering
\includegraphics[angle=0,width=0.95\columnwidth]{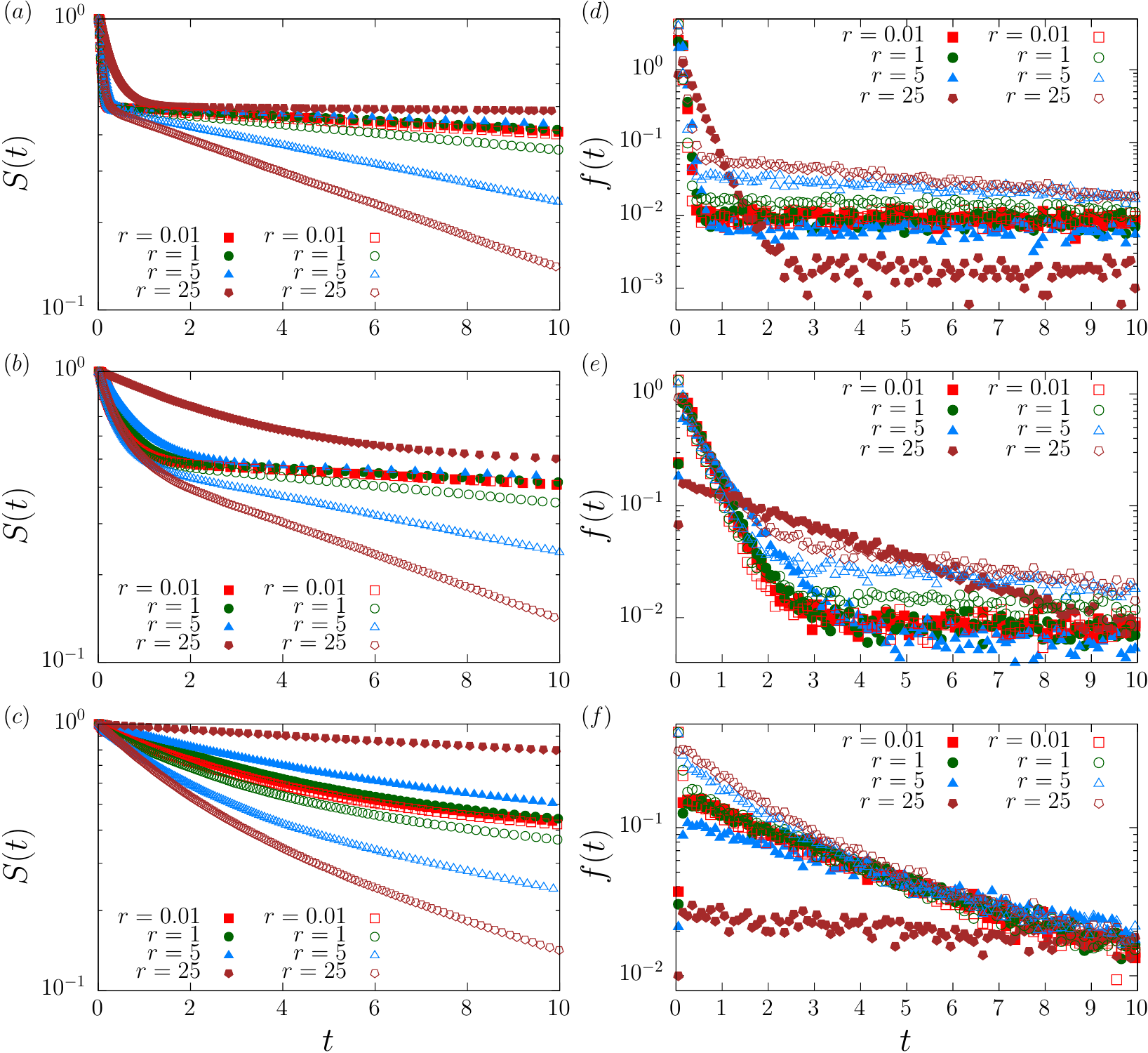} 
\caption{Survival probabilities  $S(t)$ (left column) and first passage time densities $f(t)$ (right column) corresponding to setups studied in Figs.~\ref{fig:ra-reset} ($\xr=0$ --- full symbols) and~\ref{fig:ra-reset-x05} ($\xr=0.5$ ---  empty symbols) with $\gamma=10^{-6}$.
Various panels correspond to different slopes of potential barrier: $H_{\pm}=\pm 8$ (top panel), $H_{\pm}=8/0$ (middle panel) and $H_{\pm}=8/4$ (bottom panel).
}
\label{fig:surv}
\end{figure}

\bdt{
Fig.~\ref{fig:surv} displays survival probabilities $S(t)$ and first passage time densities $f(t)$
corresponding to setups studied in Figs.~\ref{fig:ra-reset} ($\xr=0$ --- full symbols) and~\ref{fig:ra-reset-x05} ($\xr=0.5$ --- empty symbols) with the switching rate $\gamma=10^{-6}$.
The survival probability
\begin{equation}
    S(t)=1-\int_0^t f(s)ds,
\end{equation}
is the probability that particle has not been absorbed until $t$, i.e., it designs the probability of finding a particle in the domain of motion.
Facilitation of the escape kinetics visible in Figs.~\ref{fig:ra-reset} and~\ref{fig:ra-reset-x05} is further corroborated in Fig.~\ref{fig:surv} as $\xr=0.5$ results in faster decay of $S(t)$.
Survival probabilities attain typical exponential tails.
Moreover, for low values of the switching rate $\gamma$, e.g., $\gamma=10^{-6}$,  two slopes are well visible \cite{bier1993matching,dybiec2002b}, see Fig.~\ref{fig:surv}(a).
The small $t$ slope is determined by the escape over a lower barrier configuration, while the large $t$ slope by the transition over a higher barrier.
The differences between slopes is especially pronounced when MFPTs over both barrier configurations are very distinct, e.g., for $H_\pm=\pm 8$.
With larger $\gamma$ the crossover vanishes because for larger switching frequencies smaller MFPT are recorded.
The decay of $S(t)$ is the fastest for $\gamma$ resulting in the minimal mean first passage time.
The further increase in $\gamma$ increase MFPT and slows down the decay of $S(t)$.
The same information can be also deduced from first passage time densities, but this time faster escape is associated with larger value of $f(t)$.
}

%
%
\section{Summary and conclusions \label{sec:summary}}

The phenomenon of stochastic resonant activation bears some similarities with the stochastic resetting.
In the \egn{SRA phenomenon} the fine-tuned barrier modulation protocol can result in the optimization of the escape time from the potential well.
Analogously \egn{to} the escape from the finite interval restricted by two absorbing boundaries for appropriately selected $\xr$ there exists an optimal resetting protocol resulting in minimal escape time.
Nevertheless, one should keep in mind that resetting to the intermediate position $\xr>0$ replaces surmounting of the potential barrier.
More precisely, a particle can \egn{ascend} a part of the barrier by resetting what in some situations can be an artificial mechanism of climbing.

The most significant difference between escape from a potential well under stochastic resetting and due to stochastic resonant activation is observed in asymptotic behavior.
In the SRA under dichotomous noise small and large $\gamma$ asymptotics are well defined, i.e., corresponding MFPTs are finite.
Very different situation is observed for escape from the fixed potential well under resetting.
For $r \to 0$ the classical noise driven escape is recorded while for $r \to \infty$ MFPT diverges as the particle is returned to a given point all the time.

Altogether, stochastic resonant activation and stochastic resetting are two optimization mechanisms different in nature.
Stochastic resetting induces fastest escape by elimination of suboptimal, \bdt{meandering and exploring distant points}  trajectories while the efficiency of SRA relies on matching of time scales.
The fastest escape under resetting can occur \bdt{if the motion is restarted to the point which is not too distant from the absorbing boundary, i.e., $x=1$.}
\bdt{In contrast, if the motion is reset to the point which is further from the boundary than some critical distance\cite{pal2019first}, the stochastic resetting does not expedite escape kinetics.}
At the same time SRA can be recorded for \bdt{such initial conditions for which stochastic resetting is not beneficial}.
Finally, due to dynamical modulation of the potential barrier, the examination of the coefficient of variation does not always correctly identify situations in which resetting speeds up the escape process.
The rise of interest in resetting mechanisms have been initiated by studies of stochastic search strategies and their efficiency. Recent advances \cite{debruyne2020optimization} apply the idea to optimization problems, where a ``cost'' is incurred whenever the particle is reset and a ``reward'' is obtained while the particle stays near the reset point. Models of that type can be, for example, used to design optimal driving level maximizing performance of a mechanical device.
Also, different variants of stochastic diffusion processes with resetting may be realized today experimentally with colloidal particles \cite{schmitt2006} in a flow chamber \cite{admon2018experimental} or optical tweezers \cite{paneru2020colloidal}. This opens the possibility of verification tests for many theoretical investigations of resetting mechanisms, their further developments and applications. To the best of our knowledge, our study is the first one examining resetting mechanism in systems with randomly changing potentials.

%
%
\section*{Acknowledgements}

This research was supported in part by PLGrid Infrastructure and by the National Science Center (Poland) grant 2018/31/N/ST2/00598.

\section*{Data availability}
The data that support the findings of this study are available from the author (BD) upon reasonable request.

%
%
\appendix*
\section{Escape from a fixed potential under stochastic resetting\label{sec:app-escape}}

The mean first passage time $\mathcal{T}(x)$ from a fixed potential under stochastic (fixed rate $r$) resetting can be obtained by the general relation with the Laplace transform $\tilde{F}(x,s)$ of the first passage time distribution $F(x,t)$ in the absence of resetting \cite{reuveni2016optimal}
\begin{equation}
    \mathcal{T}(x) = \frac{1-\tilde{F}(x,r)}{r \tilde{F}(x,r)}.
\end{equation}
Therefore, in order to derive a formula for the MFPT under resetting it is enough to solve the resetting-free model.

The motion of a particle in the fixed potential described by the Langevin equation
    \begin{equation}
  \frac{dx}{dt}=-V'(x,t)+\sqrt{2T}\xi(t),
  \label{eq:langevin-app}
\end{equation}
is associated \cite{risken1996fokker,gardiner2009} with the Smoluchowski-Fokker-Planck  equation
\begin{equation}
\frac{\partial }{\partial t} p(x,t)= \left[ \frac{\partial}{\partial x} V'(x) + T \frac{\partial^2}{\partial x^2}  \right]p(x,t),
\label{eq:fpe}
\end{equation}
where $p(x,t)$ is the probability of finding a particle in the vicinity of $x$ at time $t$.
Similarly like in Ref.~\onlinecite{singh2020resetting} we use $V(x)= H |x|$ but in contrast to Ref.~\onlinecite{singh2020resetting} it is assumed the motion is restricted by two absorbing boundaries placed at $\pm L$.
The boundary condition for Eq.~(\ref{eq:fpe}) reads $p(x=\pm L,t)=0$.

The survival probability $q(x,t)$ \cite{goelrichter1974,risken1996fokker,gardiner2009} satisfies the backward Smoluchowski-Fokker-Planck equation
\begin{equation}
\frac{\partial }{\partial t} q(x,t)= \left[ -V'(x) \frac{\partial}{\partial x}  + T \frac{\partial^2}{\partial x^2}  \right]q(x,t),
\label{eq:bfpe}
\end{equation}
The Laplace transform of the survival probability can be found by extending the considerations carried out in Ref.~\onlinecite{singh2020resetting}
\begin{equation}
    \tilde{q}(x,s)= b_-\exp(m_-|x|)+b_+\exp(m_+|x|)+\frac{1}{s},
\end{equation}
where
\begin{equation}
m_\pm=\frac{H}{2T}\left[1\pm \sqrt{1+\frac{4sT}{H^2}}\right]    .
\end{equation}
The survival probability and its derivative are continuous at $x=0$.
Moreover, it satisfies the boundary condition $q(x=\pm L,t)=0$.
The Laplace transform of the first passage time density is given by
\begin{equation}
    \tilde{F}(x,s)=1-s\tilde{q}(x,s).
\end{equation}
After finding $b_\pm$ and $\tilde{q}(x,s)$ one gets the closed formula for the mean first passage time
\begin{widetext}
\begin{equation}
\mathcal{T}(x) = \frac{1-\tilde{F}(x,s)}{r \tilde{F}(x,s)} \bigg|_{s=r} =    \frac{ \left(1-A\right) \exp \left(\frac{H \left(L \left(A-1\right)+x A+x\right)}{2 T}\right)+\left(A-1\right) e^{\frac{H L A}{T}}-\left(A+1\right) e^{\frac{H (L-x) \left(A-1\right)}{2 T}}+A+1}{r \left(A+\left(A-1\right) e^{\frac{H x A}{T}}+1\right) e^{\frac{H (L-x) \left(A-1\right)}{2 T}}},
\label{eq:mfpt-pot-reset}
\end{equation}
\end{widetext}
where
\begin{equation}
A=\sqrt{\frac{4 r  T}{H^2}+1}    
\end{equation}

Equation~(\ref{eq:mfpt-pot-reset}) can be used to calculate the mean first passage time in the situation when the left absorbing boundary is replaced by the reflecting boundary placed at $x=0$.
Such an equivalence is recorded for processes with continuous trajectories, e.g., driven by the Gaussian white noise, see Refs.~\onlinecite{zoia2007,dybiec2017levy,gardiner2009}, and symmetric potential $V(x)$.
Therefore, Eq.~(\ref{eq:mfpt-pot-reset}) can be used to find low and large $\gamma$ asymptotics in the model of  \bdt{(stochastic)} resonant activation with the dichotomously modulated linear slope under stochastic resetting, see Figs.~\ref{fig:ra-reset} and~\ref{fig:ra-reset-x05}.

%
%

\section*{References}

\def\url#1{}

\end{document}